 \documentclass{ws-ijmpa}

\usepackage[super,compress]{cite}
\usepackage{graphicx}
\newcommand{\bea}{\begin{eqnarray}}
\newcommand{\eea}{\end{eqnarray}}

\begin{document}
\markboth{Carrington, Frey, Meggison}{Phase transitions in anisotropic graphene}

%
\catchline{}{}{}{}{}
%

\title{Phase transitions in anisotropic graphene}

\author{M.E. Carrington}
\address{Department of Physics, Brandon University\\ Brandon, Manitoba, R7A 6A9 Canada\\
and\\
Department of Physics \& Astronomy, University of Manitoba\\ Winnipeg, Manitoba, R3T 2N2 Canada}

\author{A.R. Frey}
\address{Department of Physics, University of Winnipeg\\ Winnipeg, Manitoba, R3M 2E9 Canada\\and\\Department of Physics \& Astronomy, University of Manitoba\\ Winnipeg, Manitoba, R3T 2N2 Canada}

\author{B.A. Meggison}
\address{Department of Physics \& Astronomy, University of Manitoba\\ Winnipeg, Manitoba, R3T 2N2 Canada}

\maketitle

\begin{history}
\received{04/01/2021}
\end{history}

\begin{abstract}
We study the effect of anisotropy on phase transitions in graphene. 
We work with an low energy effective field theory which is strongly coupled, and solve a coupled set of Schwinger-Dyson equations. 
We show that the effect of anisotropy is to reduce the critical coupling. 
\end{abstract}

\ccode{PACS numbers:}


\section{Introduction}

Graphene has been the studied by physicists for many years. 
In part, the motivation is technological applications, which include the development of graphene based electronic devices. 
In this context, the ability to produce a phase transition between the semi-metal and insulating states is important. 
The system is also interesting to theoretical physicists for a number of reasons. 
Close to the phase transition, graphene can be described using a low energy effective field theory which is commonly called reduced QED, or pseudo QED \cite{marino,miransky}. 
The effective coupling constant is large enough that non-perturbative methods must be used.
In addition, both reduced QED and QCD exhibit chiral symmetry breaking. 
Thus graphene has some of the interesting features of QCD without all of its complexity. 

We are interested in the effect of anisotropy on the semi-metal/insulating phase transition in graphene. 
Physically, anisotropy could be introduced as physical strain on the graphene lattice, or possibly using an applied magnetic field. 
We will include anisotropy in our calculation at the level of the effective Lagrangian, by considering a Fermi velocity which is not isotropic in space.
There are several previous calculations in the literature that are similar in their approach \cite{sharma1, sharma2, xiao1}  
but used  numerous restrictive assumptions to reduce the difficulty of the numerical calculations. 
The effect of these approximations is, however, difficult to predict, and different approximations have led to different predictions about the direction that the critical coupling in an anisotropic system moves, relative to the isotropic one. 
In a previous paper \cite{Carrington:2020qfz}, we have presented results obtained numerically from a coupled set of Schwinger-Dyson (SD) equations using only two approximations: we truncated the hierarchy of SD integral equations using a common vertex ansatz (which will be discussed below), and we used a 1-loop approximation for the photon polarization tensor. In this paper, we relax the second of these approximations. We solve the complete set of back-coupled SD equations for the  non-perturbative dressing functions of the fermion and photon propagators. 
We find that back-coupling reduces both the critical coupling and the effect of anisotropy.

\section{Notation and Setup}

We use throughout natural units ($\hbar=c=1$) and work in Euclidean space. We use capital letters and Greek indices for (2+1)-dimensional vectors: for example $P_\mu = (p_0,p_1,p_2) = (p_0,\vec p)$ and $P^2=p_0^2+p^2$. 
We define $Q=K-P$ where $P$ is the external momentum and $K$ is the loop momentum. We write $dK = \int dk_0\,d^2k/ (2\pi)^3$ and similarly for other integration variables. 

The Euclidean action of the low energy effective theory is 
\begin{equation}
\label{action}
S=\int d^3 x \sum_{a}\bar\psi_a \left(i\partial_\mu -e A_\mu\right)M_{\mu\nu}\gamma_\nu \psi _a - \frac{\epsilon}{4e^2}\int d^3x F_{\mu\nu}\frac{1}{2\sqrt{-\partial^2}}F_{\mu\nu} \text{ + gauge fixing}.
\end{equation}
We work in covariant gauge, and the Feynman rules for the bare theory are
\bea
\label{bareFR}
&& S^{(0)}(P) = -\big[i\gamma_\mu M_{\mu\nu} P_\nu\big]^{-1}\,\\[2mm]
&& G^{(0)}_{\mu\nu}(P)=\big[\delta_{\mu\nu}-\frac{P_\mu P_\nu}{P^2}(1-\xi)\big]\,\frac{1}{2\sqrt{P^2}}\, \\[1mm]
\label{barevert}
&& \Gamma^{(0)}_\mu = M_{\mu\nu}\gamma_\nu\,
\eea
where we have defined
\bea
\label{Mdef}
M = 
\left[\begin{array}{ccc}
~1~ & ~0~ & ~0~ \\
0 & v_{1}  & 0 \\
0 & 0 & v_{2}   \\
\end{array}
\right]\,
\eea
and the three four-dimensional $\gamma$-matrices  satisfy $\{\gamma_\mu,\gamma_\nu\} = 2\delta_{\mu\nu}$.
In the isotropic limit, $v_1=v_2=v_F\equiv c/300$. In our calculation the Fermi velocity is the geometric mean $v_F=\sqrt{v_1v_2}$,
and the anisotropy parameter is the ratio $\eta=v_1/v_2$.

The Landau gauge non-perturbative photon propagator is 
\bea
\label{fullG}
&& G_{\mu\nu} =  
\big(\delta_{\mu\nu}-\frac{P_\mu P_\nu}{P^2}\big)\frac{1}{G_T(p_0,\vec p)}+ \big(\delta_{\mu 0}-\frac{p_0P_\mu}{P^2}\big)\big(\delta_{\nu 0}-\frac{p_0P_\nu}{P^2}\big)\left(\frac{1}{G_L(p_0,\vec p)}-\frac{1}{G_T(p_0,\vec p)}\right)\, ,\nonumber\\[4mm]
\label{fullPi}
\eea
with
\bea
&& G_T(p_0,\vec p) = 2\sqrt{P^2}+\alpha(p_0,p)\, ,\nonumber\\
&& G_L(p_0,\vec p) = 2\sqrt{P^2}+\alpha(p_0,p)+\gamma(p_0,p) \nonumber\\
&& {\rm Tr}\Pi(p_0,p) = \big(2\alpha(p_0,p) + \gamma(p_0,p)\big)\, \nonumber\\
\label{00L}
&& \Pi_{00}(p_0,p) = \frac{p^2}{P^2}\big(\alpha(p_0,p) + \gamma(p_0,p)\big)\,. 
\eea
The functions $\alpha(p_0,p)$ and $\gamma(p_0,p)$ are photon dressing functions, whose definitions are given in the last two lines of equation (\ref{00L}). 
The dressed fermion propagator depends on four independent dressing functions which we denote $Z_p\equiv Z(p_0,\vec p)$, $A_{\rm 1p}\equiv A_1(p_0,\vec p)$,  $A_{\rm 2p}\equiv A_2(p_0,\vec p)$ and $D_p\equiv D(p_0,\vec p)$. The propagator is
\bea
S = \frac{1}{S_p}\left[ i \gamma_\mu M_{\mu\alpha} F(p_0,\vec p)_{\alpha\nu} P_\nu +D_p \right]
\eea
with
\bea
F(p_0,\vec p) = \left[
\begin{array}{ccc}
 Z_p & 0 & 0 \\
 0 & A_{\text{1p}} & A_{\text{2p}} \\
 0 & -A_{\text{2p}} & A_{\text{1p}} \\
\end{array}
\right]\,\label{F-def}
\eea
\bea
S_p = p_0^2 Z_p^2 + v_1^2 \left(p_1 A_{\text{1p}}+p_2 A_{\text{2p}}\right){}^2 
+ v_2^2 \left(p_2 A_{\text{1p}} - p_1 A_{\text{2p}}\right){}^2+D_p^2 \,.\label{Sp}  
\eea
The bare fermion propagator is obtained by setting $Z(p_0,\vec p) = A_1(p_0,\vec p) =1$ and $A_2(p_0,\vec p)= D(p_0,\vec p)=0$.

The SD equations for the fermion self energy and photon polarization are 
\bea
\label{fermion-SD}
&& \Sigma(p_0,\vec p) = e^2\int dK \,G_{\mu\nu}(q_0,\vec q)\,M_{\mu\tau}\,\gamma_\tau \, S(k_0,\vec k) \,\Gamma_\nu\, \\
\label{photon-SD}
&& \Pi_{\mu\nu}(p_0,\vec p) = -e^2\int dK \,{\rm Tr}\,\big[S(q_0,\vec q) \, M_{\mu\tau} \, \gamma_\tau \, S(k_0,\vec k)\, \Gamma_\nu\big]\,.
\eea
To leading order in $(v_1/c,v_2/c)$ the only component of the propagator (\ref{fullG}) that contributes to the fermion self energy $\Sigma$ is the piece $G_L$, so we only need to calculate the zero-zero component of the polarization tensor (see equation (\ref{00L})). 
To truncate the SD hierarchy we use an ansatz for the 3-vertex denoted $\Gamma$ in equations (\ref{fermion-SD}, \ref{photon-SD}). 
In the isotropic theory one commonly uses the Ball-Chiu vertex ansatz \cite{ball-chiu-1,ball-chiu-2}, which is constructed to preserve gauge invariance. 
We will use a modified version of this vertex which is suitable for an anisotropic theory, and furthermore we will use a truncation that is numerically more stable. It is known that the effects of this truncation are small in the isotropic theory \cite{mec1}, and a check for the anisotropic theory is currently in progress. We use 
\bea
\Gamma_\mu(P,K) && = \frac{1}{2}\big[F(p_0,\vec p)_{\mu\alpha}^T+F(k_0,\vec k)_{\mu\alpha}^T\big]M_{\alpha\beta}\gamma_\beta  
 \label{ballchiu} 
\eea
where $(P,K)$ are the momenta of the incoming and outgoing fermions, respectively.

The SD equations for the fermion dressing functions and the zeroth component of the polarization tensor are obtained by taking the appropriate projections of (\ref{fermion-SD}) and (\ref{photon-SD}). 
The results are below:
\bea
\label{Znew}
&& Z(p_0,\vec p) = 1-\frac{2\alpha \pi v_F}{p_0}\int \frac{dK}{Q^2 S_k G_L} \,k_0 q^2 Z_k \left(Z_k+Z_p\right)\, , \\
&& A_1(p_0,\vec p) = 1 + \frac{2\alpha \pi v_F}{p^2}\int  \frac{dK}{Q^2 S_k G_L}
\bigg[ k_0 q_0 Z_k (\vec p\cdot\vec q)  \left(A_{\text{1k}} +A_{\text{1p}}+Z_k+Z_p\right) \nonumber \\
&& ~~~~    +q^2 A_{\text{1k}} \left(Z_k+Z_p\right) (\vec k\cdot\vec p)  
    + k_0 q_0 Z_k \left(A_{\text{2k}} + A_{\text{2p}}\right) (\vec p\times \vec q) 
     - q^2 A_{\text{2k}} \left(Z_k+Z_p\right) (\vec k\times \vec p)  \bigg]\,  \nonumber \\ 
\label{A1new}\\
&& A_2(p_0,\vec p) =  \frac{2\alpha \pi v_F}{p^2}\int \frac{dK}{Q^2 S_k G_L}
\bigg[-k_0 q_0 Z_k  (\vec p \times \vec q)  \left(A_{\text{1k}}+A_{\text{1p}}+Z_k+Z_p\right) \nonumber \\
&& ~~~~  +q^2 A_{\text{1k}} \left(Z_k+Z_p\right) (\vec k\times \vec p)  
   + k_0 q_0 Z_k \left(A_{\text{2k}}+A_{\text{2p}}\right) (\vec p\cdot\vec q)  
   + q^2 A_{\text{2k}} \left(Z_k+Z_p\right)  (\vec k\cdot \vec p)  \bigg]\, \nonumber \\ 
\label{A2new}   \\
&& \label{Dnew} D(p_0,\vec p) = 2\alpha \pi v_F\int \frac{dK}{Q^2 S_k G_L}   \, q^2 D_k \left(Z_k+Z_p\right)\,  \\
&& \Pi_{00}(p_0,p) = -16 \pi v_F \alpha \int \frac{dK}{S_k S_q} \bigg[
\left(Z_k+Z_q\right) \left(D_k D_q-k_0 q_0 Z_k Z_q\right)
+ A_{\text{1k}} A_{\text{2q}} \left(Z_k+Z_q\right) (\vec k\times \vec q)_v \nonumber \\
&&~~~~ + A_{\text{1q}} A_{\text{2k}} \left(Z_k+Z_q\right) (\vec q\times \vec k)_v 
+ A_{\text{1k}} A_{\text{1q}} \left(Z_k+Z_q\right) (\vec k\cdot \vec q)_{v}  
+ A_{\text{2k}} A_{\text{2q}} \left(Z_k+Z_q\right) (\vec k\cdot \vec q)_{v'}  \bigg]\, , \nonumber\\ \label{Lnew}
\eea
where we have used the notation
$\alpha = e^2/(4\pi \epsilon v_F)$, $\vec k \cdot \vec p = k_1 p_1 + k_2 p_2$, $(\vec k \cdot \vec p)_v = v_1^2 k_1 p_1 + v_2^2 k_2 p_2$, $(\vec k \cdot \vec p)_{v'} = v_2^2 k_1 p_1 + v_1^2 k_2 p_2$, $\vec k \times \vec p = k_1 p_2 - k_2 p_1$, and $(\vec k \times \vec p)_v = v_1^2 k_1 p_2 - v_2^2 k_2 p_1$. 
We find numerical solutions of the SD equations that have the symmetries of the bare theory. The dressing functions $Z$, $A_1$, and $D$ are assumed even under the transformations $p_0\to -p_0$, $p_1\to -p_1$, and $p_2\to -p_2$, and even under the interchange $(p_1,v_1) \leftrightarrow (p_2,v_2)$.
The function $A_2$ is even under $p_0\to -p_0$ and odd under all the other transformations above. It is easy to see that the equations (\ref{Znew}-\ref{Lnew}) respect these symmetries. 

In a previous paper \cite{Carrington:2020qfz} we used a 1-loop result for the photon polarization, instead of solving the coupled self-consistent equation (\ref{Lnew}). 
This approximation 
is based on the fact that the the fermion density of states vanishes at the Dirac points, and is widely adopted because of the significant numerical advantages. 
The integrals in equation (\ref{Lnew}) can be calculated directly when bare fermion propagators are used, and give the simple analytic result
\bea
\label{Piana}
\Pi^{\rm 1\,loop}_{00}(p_0,p) = \frac{\pi\alpha}{\sqrt{v_{1} v_{2}}}
\,\frac{p_1^2 v_{1}^2 + p_2^2 v_{2}^2}
{\sqrt{p_0^2+p_1^2 v_{1}^2 + p_2^2 v_{2}^2}} \,.
\eea
In this paper we solve the full set of coupled equations (\ref{Znew}-\ref{Lnew}) and compare with the result obtained using (\ref{Piana}) instead of (\ref{Lnew}).

\section{Numerics and Results}	

Reduced QED is finite except for a divergence in the photon polarization tensor, which can be removed by a subtraction. We use $\Pi^R_{\mu\nu}(P) = \Pi_{\mu\nu}(P)-\Pi_{\mu\nu}(0)$ and in the rest of this paper we suppress the superscript $R$.
The momentum variables are discretized using a logarithmic scale to increase the number of grid points close to the origin, where the dressing functions vary the most and accuracy is most important. 
The dressing functions themselves are fairly smooth but the integral equations involve integrable singularities, which means that a larger number of grid points are needed for discretized integration variables. 
We use Gauss-Legendre quadrature which is numerically more efficient than a constant partitioning integration procedure. 
Interpolations are done using three-dimensional linear interpolation.
Our calculations were done using $1.64\times 10^{4}$ external grid points and $3.2\times 10^{5}$ internal grid points, and we have checked that our results are very stable when the number of external and/or internal grid points is increased. 
Another numerical issue is that the number of iterations required to converge to a solution of the SD equations increases significantly as $\alpha$ approaches the critical value, due to what is known as `critical slowing down.' 
To circumvent this problem, the iteration procedure is initialized for a new value of the coupling, from the closest value that has previously been calculated.
The smallest values of $\alpha$ for which we have obtained solutions require about 600 iterations to converge. 
We also note that convergence is slower when the anisotropy of the system increases because the dressing function $A_2$ (which is zero in the isotropic calculation) becomes increasingly important as the anisotropy increases. 
All of our calculations are done in fortran, and parallelized using openMPI 4.0.1.

\begin{figure}[h]
\begin{center}
\includegraphics[width=0.95\linewidth]{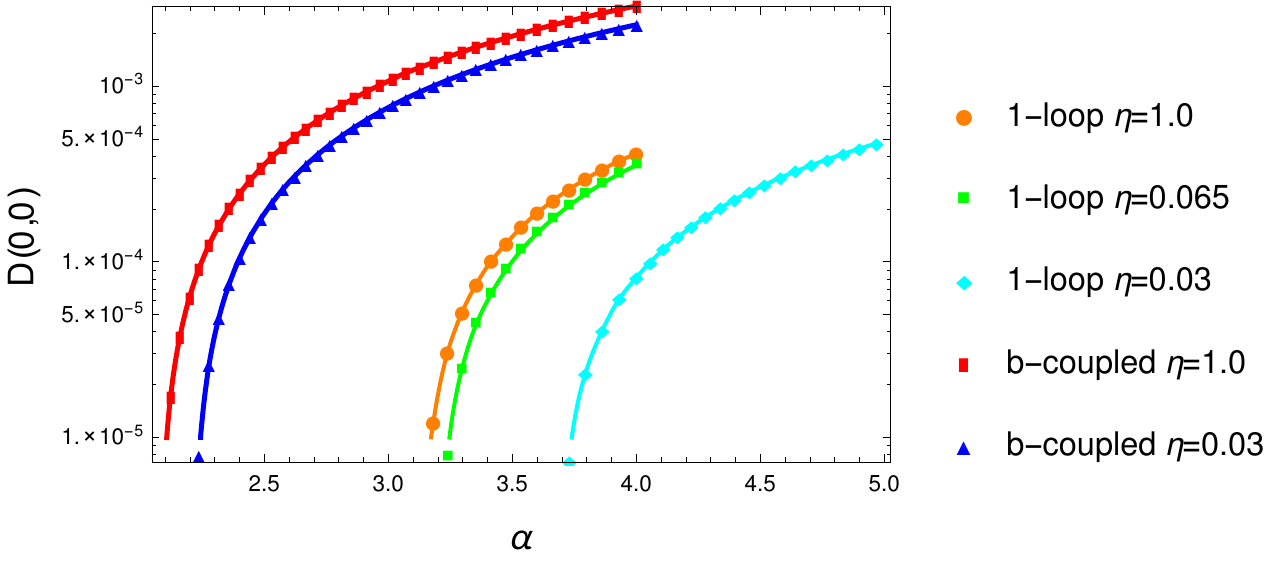}
\end{center}
\caption{The condensate $D(0,0)$ vs. coupling for different values of the anisotropy parameter from the full back-couopled calculation and using the 1-loop approximation to the polarization tensor. }
\label{condensate}
\end{figure}

In Fig.~\ref{condensate} we show the value of the condensate $D(0,0)$ versus coupling for different values of the anisotropy parameter from the full back-coupled calculation, and using the 1-loop approximation for the photon polarization tensor. The curves for the back-coupled calculation with $\eta=1.0$ and $\eta=0.65$ are almost on top of each other, and therefore the $\eta=0.65$ line is not shown on the graph. 

We calculate the critical coupling by extrapolating to find the value of the coupling for which the condensate would be zero. 
We fit the data using a polynomial fit with polynomials of degree 3 to 5, a Hermite polynomial fit working to orders 3 to 5, and a cubic spline fit, and in all cases the differences between any two fits is smaller than the quoted uncertainty, which indicates the extrapolation does not introduce any sizeable error.  
Our numerical results for the critical coupling are shown in the first and third columns of Table \ref{criticalcoupling}. 
We have estimated the uncertainty by comparing the results of the extrapolation procedure using our full data array, and the array with the smallest calculated point removed.
The second and fourth columns show the isotropic result obtained using a similar method in Refs. \cite{mec2, mec1}. The fifth column shows the results from Ref. \cite{xiao1}, taking into account that the definition of $\eta$ in that paper is equivalent to $1/\eta$ in ours. The numbers quoted  are estimated from their Fig. 7 and are only approximate. 
\begin{table}[ph]
\tbl{Summary of critical coupling results.}
{\begin{tabular}{cc c c c c c@{}} \toprule
$\eta$ & $\alpha_c$ b-coupled & $\alpha_c$ b-coupled \cite{mec2} & $\alpha_c$ 1-loop & ~~ $\alpha_c$ 1-loop \cite{mec1} ~~ & ~~ $\alpha_c$ \cite{xiao1} ~~\\ 
\hline1 & 2.08 $\pm$ 0.004& 2.06 $\pm$ 0.01 & 3.12 $\pm$  0.02 & 3.12 $\pm$ 0.01 & $\approx$ 0.92  \\
0.65 &2.10 $\pm$ 0.006  & & 3.21 $\pm$ 0.02 &  & $\approx 0.94$  \\
0.3 & 2.20 $\pm$ 0.02~ & & 3.70 $\pm$ 0.04 &  & $\approx 1.05$  \\ \botrule
\end{tabular} \label{criticalcoupling}}
\end{table}
The results in Table \ref{criticalcoupling} show that the introduction of anisotropy increases the critical coupling. 
In the back-coupled calculation, the critical coupling is smaller, and the effect of anisotropy is reduced. 
We also see that the effect of anisotropy is smaller than the change obtained by replacing the 1-loop photon polarization by the full back coupled result. This conclusion may be related to the fact that the Coleman-Hill theorem is valid for reduced-QED \cite{Dudal:2018mms}, which means that parity odd radiative corrections to the photon polarization should exist only up to one loop.

\section{Conclusions}

We have studied the effect of anisotropy on the semi-metal insulatar phase transition in graphene using a low energy effective theory. 
We use a Schwinger-Dyson method, and truncate the hierarchy of integral equations using a 
Ball-Chiu-like vertex ansatz. Full frequency dependence of the dressing functions is included. 
We have done a fully back-coupled calculation in which the photon dressing function is determined self-consistently, and compared with results obtained using a 1-loop  photon polarization tensor, which is a commonly used approximation in the literature. 
Our 1-loop results show that the effect of anisotropy is greater than predicted
by previous calculations, and that it  increases the critical coupling. 
When back-coupling is included, both the critical coupling and the effect of anisotropy is reduced. 

We also remind the reader that the value of the critical coupling produced by any calculation based on an effective theory is not expected to be exact, since there are screening effects that could be important and are not included. The goal of the calculation is to establish whether or not anisotropy could reduce the critical coupling, and therefore make it easier to produce an insulating state. 
Our results indicate anisotropy increases the critical coupling, instead of moving it downward toward values that would be physically realizable.

\section*{Acknowledgments}
This work has been supported by the Natural Sciences and
Engineering Research Council of Canada Discovery Grant program.
This research was enabled in part by support provided by WestGrid
(www.westgrid.ca) and Compute Canada Calcul Canada (www.computecanada.ca).



\begin{thebibliography}{0}    

\bibitem{marino}
Marino, E.C., Nuclear Physics B 408, 551 (1993).
\bibitem{miransky}
Gorbar, E.V. and Gusynin, V.P. and Miransky, V.A. and Shovkovy, I.A.
 Phys. Rev. B66, 045108, (2002).
 
\bibitem{sharma1}
Anand Sharma and Valeri N. Kotov and Antonio H. Castro Neto,
(2012), arXiv:1206.5427 [cond-mat.str-
el].

\bibitem{sharma2}
Sharma, Anand and Kotov, Valeri N. and Castro Neto, Antonio H.,
    Phys. Rev. B,
95,
(2017).

\bibitem{xiao1}
Xiao, Hai-Xiao and Wang, Jing-Rong and Feng, Hong-Tao and Yin, Pei-Lin and Zong, Hong-Shi.
    Phys. Rev. B,
155114,
(2017).
 
\bibitem{Carrington:2020qfz}
M.~E.~Carrington, A.~R.~Frey and B.~A.~Meggison,
Phys. Rev. B 102, 125427 (2020).

\bibitem{ball-chiu-1}
Ball, James S. and Chiu, Ting-Wai,
Phys. Rev. D,
2542, (1980).


\bibitem{ball-chiu-2}
Ball, James S. and Chiu, Ting-Wai, 
Phys. Rev. D,
2550,
(1980).

\bibitem{mec1} Carrington, M.E. and Fischer, C.S. and von Smekal, L. and Thoma, M.H., Phys. Rev. B,
125102,
(2016).

\bibitem{mec2}
Carrington, M.E. and Fischer, C.S. and von Smekal, L. and Thoma, M.H.,
    Phys. Rev. B, 115411, (2018).
    
\bibitem{Dudal:2018mms}
D.~Dudal, A.~J.~Mizher and P.~Pais,
Phys. Rev. D \textbf{98}, 065008 (2018).



\end{thebibliography}
\end{document}